\documentclass[12pt]{revtex4}
\usepackage{epsfig}
\begin{document}
\pagestyle{empty}

\title{
 {\bf Residual correlations between decay products
 of $\pi^0\pi^0$ and $p\Sigma^0$ systems}}

\author{
 A.Stavinskiy$^a$, K.Mikhailov$^a$, B.Erazmus$^b$,
 R.Lednicky$^{c,d}$}

\affiliation{$^a$Institute of Theoretical and Experimental Physics,
B.~Cheremushkinskaya 25, 117259, Moscow, Russia\\
$^b$SUBATECH, UMR Univ., EMN, IN2P3/CNRS,
4 rue A.Kastler,44307 Nantes, France\\
$^c$Joint Institute for Nuclear
Research, Dubna, Moscow Region, 141980, Russia\\
$^d$Institute of Physics ASCR,
Na Slovance 2, 18221 Prague 8, Czech Republic
}




\begin{abstract}
Residual correlations between decay products due to
a combination of both correlations
between parents at small relative velocities and
small decay momenta are discussed.
Residual correlations between photons from pion
decays are considered as a new possible source of information on
direct photon fraction. Residual correlations
in $p\gamma$ and $p\Lambda$ systems due to $p\Sigma^0$ interaction
in final state are predicted based on the $p\Sigma^0$ low
energy scattering parameters deduced from
the spin-flavour SU$_6$ model by Fujiwara et al.
including effective meson exchange potentials and
explicit flavour symmetry breaking
to reproduce the properties of the two-nucleon system and the
low-energy hyperon-nucleon cross section data.
The $p\gamma_{\Sigma^0}$ residual correlation is concentrated
at $k^* \approx 70$ Mev/$c$ and its shape and intensity
appears to be sensitive to
the scattering parameters and space-time dimensions of the source.
The $p\Lambda_{\Sigma^0}$ residual correlation
recovers the negative parent $p\Sigma^0$ correlation for
$k^* > 70$ Mev/$c$. The neglect of this negative
residual correlation would lead to the underestimation of
the parent $p\Lambda$ correlation effect and
to an overestimation of the source size.
\end{abstract}

\maketitle

\newpage
\pagestyle{plain}
\section{Introduction}

Due to the effects of quantum
statistics (QS) and final state interaction (FSI),
the momentum correlations of two or more particles at small
relative velocities, i.e. at small relative momenta in their
center-of-mass (c.m.) system,
are sensitive to the space-time characteristics of the
production processes on a level of fm $=10^{-15}$ m.
Consequently, these correlations are widely used
as a correlation femtoscopy tool
providing a unique information on the reaction mechanism
which is hardly accessible by other means
(see, e.g., recent reviews \cite{led04,lis05}).

The momentum correlations of two particles with
four-momenta $p_1$ and $p_2$ are studied with the help of
the correlation function ${\mathcal R}(p_{1},p_{2})$
which is usually defined as the
ratio of the measured distribution of the three-momenta
of the two particles to the
reference one obtained by mixing particles from different
events of a given class,
normalized to unity at sufficiently
large relative momenta.
It can be also written as the ratio of the two-particle
production cross section to the product of the
single-particle ones
\begin{equation}
\label{eq:corfundef}
 {\cal R}({\bf p}_1,{\bf p}_2) = N
\frac{d^6\sigma / d^3{\bf p}_1 d^3{\bf p}_2}
     {d^3\sigma / d^3{\bf p}_1            \cdot
      d^3\sigma / d^3{\bf p}_2           },
\end{equation}
where $N$ is the normalization factor which is sometimes
taken weakly dependent on the relative momentum
to account for the effect of possible non-femtoscopic
correlations.

At a first glance, one can hardly expect any
correlations between photons from the decays of
different neutral pions.
The spatial separation between such photons is of the order
of $10^6$ fm and the corresponding Bose-Einstein enhancement
is extremely narrow and practically unobservable.
Nevertheless,
due to femtoscopic QS correlations
between parent pions at small relative momenta as well as
due to a small decay momentum,
correlations between decay photons from different neutral pions
should exist and have been experimentally observed
\cite{residual_gg,wa98}.
The small decay momentum guarantees that a small
relative momentum between photons corresponds to
a small relative momentum between parent pions.
For this kinematic reason the QS correlation between neutral pions
is transferred to decay photons, being however smeared
due to randomly distributed directions of the
decay three-momenta in the respective parent rest frames.
We shall refer to such correlations as the residual
ones (see also~\cite{residual_gg,wa98,residual_BB}).

The residual correlations are important not
only for two-photon system.
For example, the two-baryon correlations are
also affected by residual correlations arising
from the FSI and QS correlations in the systems
involving strange baryons since the
decay momenta in their decays
(e.g., $\Lambda \rightarrow p \pi^{-}$ or
 $\Sigma^0 \rightarrow \Lambda \gamma$)
are not so large to destroy
the original correlations \cite{residual_BB}.
The relative importance of the residual correlations
is however quite different for two-photon and
two-baryon systems. Usually, the residual two-photon
correlation almost completely dominates over the
correlation of direct photons.
The residual correlations in two-baryon and other systems
become more and more important
with the increasing collision energy due to
increasing fraction of the produced strange particles.
Their analysis is therefore an up-to-date task.

The residual correlations do not represent only the
distorting effect which introduces additional systematic errors
in correlation studies.
We would like to pay attention to the fact
(to our knowledge, for the first time)
that the residual correlation itself is a valuable
source of femtoscopic information.

\section{Two-photon correlations}

To study the residual correlations between photons
($\gamma_{\pi^{0}}$)
from neutral pion decays ($\pi^0\to \gamma\gamma$),
we have assumed that the
photons are produced either through these decays
or "directly" (similar to the production of $\pi^0$'s)
and generated neutral pions and direct photons
($\gamma_{\rm D}$)
according to the same thermal-like momentum distribution
\begin{equation}
\label{eq:momentum}
 dN/dp \propto (p^2/E) exp(-E/T_0),
\end{equation}
where $ T_0 $ =168 MeV.
Usually, there is the experimental threshold
in the energy of detected photons.
To take it into account
we have rejected photons with $E_{\gamma} < 80$ MeV.
Such a cut increases the relative strength of the
correlation of direct photons
and modifies the relative strength and shape
of residual correlations.

The fraction $d=N(\gamma_{\rm D})/[N(\gamma_{\pi^{0}})+
N(\gamma_{\rm D})] $ of direct photons
was considered as a simulation parameter  ($ d=0, 5, 10, 20 \% $).
Only the QS correlation has been introduced by giving
each simulated pair of neutral pions and direct (unpolarized)
photons a weight (see, e.g., \cite{ll1})
\begin{eqnarray}
\label{eq:QSweight}
 {\cal R}({\bf p}_1,{\bf p}_2)&=&1+\lambda
 \langle\cos(2{\bf k}^*{\bf r}^*)\rangle
\nonumber \\
 &=&
 1+\lambda\exp(-Q^{2}r_0^2),
\end{eqnarray}
where ${\bf Q}=2{\bf k}^*$ and ${\bf r}^*$ are respectively the relative
three-momentum of the two particles and spatial separation vector
of their emitters in the two-particle rest frame;
$Q=\sqrt{-(p_1-p_2)^2}$ for the considered equal-mass particles.
The correlation strength parameter
$\lambda=1$ for pions and 1/2 for unpolarized photons.
A Gaussian ${\bf r^*}$-distribution
with the same radius parameter $r_{0}$ has been assumed
in the averaging in Eq. (\ref{eq:QSweight})
over the spatial separation
${\bf r^*}$ for both pion and photon emitters:
\begin{equation}
\label{eq:space}
 d^3 N/d^3 {\bf r^*} \sim \exp(-{\bf r}^{*2}/4r^2_0).
\end{equation}
This is a reasonable assumption for pions emitted with
moderate transverse momenta. However, for photons, it
leads to the non-realistic dependence of the correlation function
on the outward component of the relative momentum
in the longitudinally co-moving system
(the component in the direction of the pair transverse momentum)
due to the diverging Lorentz factor of the transformation
to the two-photon rest frame at $Q\to 0$.

\begin{figure}[ht]
  \begin{center}
    \includegraphics[width=16cm]{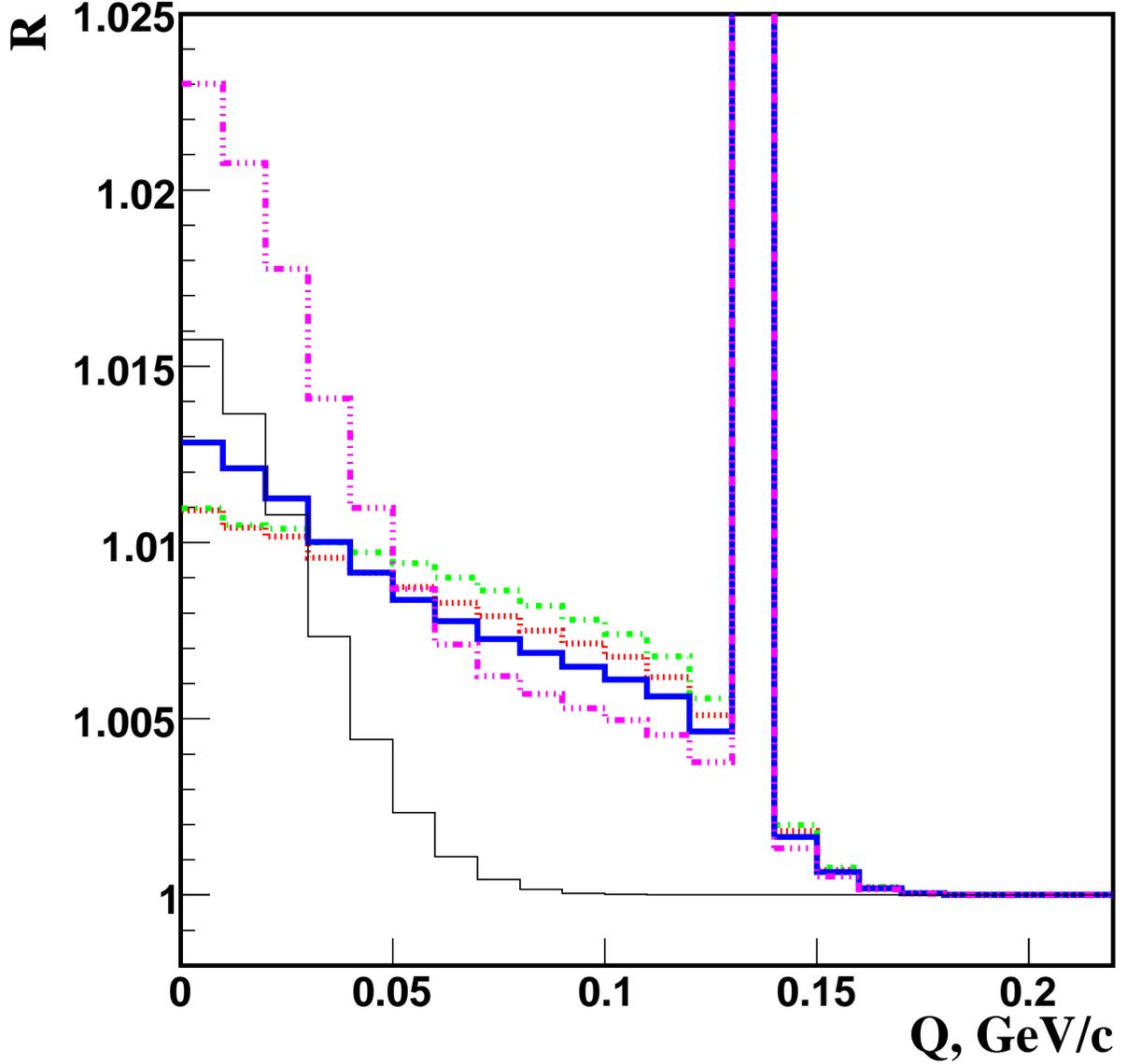}
  \end{center}
  \caption{The two-photon correlation functions
  calculated with the source size parameter $r_0=5$ fm
  and different direct photon fractions
  $d=N(\gamma_{\rm D})/[N(\gamma_{\pi^{0}})+
     N(\gamma_{\rm D})] $.
  The histograms correspond to $d=0$ (dashed-dotted),
  0.05 (dotted), 0.10 (solid), 0.20 (dashed-triple-dotted).
  The thin solid histogram corresponds to $d=0.20$ and the
  residual correlation switched off.}
  \label{fig:photons}
\end{figure}

The two photon correlation functions corresponding to
the radius parameter $r_{0} = 5$ fm and different
direct photon fractions $d$ are shown in
Fig.~\ref{fig:photons} as functions of the
relative momentum $Q=2k^*$ in the two-photon c.m. system
assuming the ideal three-momentum resolution.
The correlation functions are
normalized to unity at $Q \gg m_{\pi}$.
The peak at $Q = m_{\pi^0}$ is related to photon pairs from
the same $\pi^0$. The width of the peak in a real experiment
depends on the three-momentum resolution.
The residual correlations between
decay photons from different  $\pi^0$'s
($\sim 81 \% $ contribution for $ d=0.1$)
result in a smooth structure
at $Q < 0.17$ GeV.
The uncorrelated background ($ \sim 18 \% $ contribution
for $ d=0.1 $) arises from
$\gamma_{\pi^0}\gamma_{\rm D}$-pairs.
The pairs of direct photons
($\sim 1\%$ contribution for $ d=0.1 $)
provide the interference enhancement with a width of
$1/r_0 \sim 40$ MeV/$c$.

It is important that
\begin{enumerate}
\item[(i)] the residual correlations represent a first order effect
in the direct photon fraction $d$, to be compared with the
second order effect of the interference
correlations of direct photons;
\item[(ii)] the residual correlation effect appears to be wider than
the interference effect for $\pi^0$'s or direct photons.
\end{enumerate}
As a result, for some combinations of a large
source size $r_0$ and a small direct photon fraction $d$, the
direct photon interference is practically unobservable while the
residual correlations can still be used to measure $d$.

It is important for the suggested method
that the residual correlation function of decay photons
(depending on the three-momentum spectrum of neutral pions,
their correlations and experimental conditions)
could be predicted with sufficient accuracy.
This is in contrast with the model-independent correlation
measurement of the direct photon fractions in the experiment
WA98 \cite{wa98} exploiting the quadratic relation between the
correlation strength parameter $\lambda$ and the direct photon
fraction (valid in sufficiently narrow interval of the three-momenta
of the selected photon pairs) and nearly constant residual
correlation function of decay photons at $Q< 90$ MeV/$c$.



\section{The residual correlations in $h\Lambda$ and $h\gamma$ systems
induced by the correlations in $h\Sigma^0$ system}

\subsection{Kinematic considerations}

Another interesting example of residual correlations are the
correlations in $h\Lambda$ and $h\gamma$ systems
(where $h$ is a hadron) induced by the
$h\Sigma^0$ correlation at small relative $h-\Sigma$
momenta $2 K$
in the $h\Sigma^0$ c.m. system.
One can express the momentum $k^*$ of the hadron $h$ in
the $h\Lambda$ or $h\gamma$ c.m. system
through the hadron three-momentum $-{\bf K}$ as
\begin{eqnarray}
\label{k_k}
&&k^*= m_h\left[\frac{z^2-m^2/m_h^2}{1+m^2/m_h^2+2z}
\right]^{1/2}
\nonumber\\
&&z=\frac{\omega_{\rm D}(\omega_h\omega_\Sigma+K^2)}
{m_\Sigma m_h^2}
+\frac{p_{\rm D}K(\omega_h+\omega_\Sigma)}
{m_\Sigma m_h^2}\zeta,
\end{eqnarray}
where $p_{\rm D}=74$ MeV/$c$ is the
decay momentum in the decay $\Sigma^0\to\Lambda\gamma$,
$m$ is the mass of the decay particle (a proton or a photon),
$\omega_{\rm D}=(m^2+p_{\rm D}^2)^{1/2}$ is the corresponding
decay energy,
$\omega_i=(m_i^2+K^2)^{1/2}$ are the energies of the
particles $i=h, \Sigma^0$ in the $h\Sigma^0$ c.m. system and
$\zeta$ is the uniformly distributed cosine of the angle
between the vectors ${\bf p}_{\rm D}$ and ${\bf K}$.

Since the decay momentum in the $\Sigma^0\to\Lambda\gamma$
is rather small,
the velocity of the decay-$\Lambda$ is
close to the parent ($\Sigma^0$) velocity
and so a substantial part of the $h\Sigma^0$ correlation
at small relative $h-\Sigma^0$ velocities
is transferred to the $h\Lambda$ correlation
at small $h-\Lambda$ relative velocities.
More quantitatively, it follows from Eq. (\ref{k_k})
that the parent correlation effect
at $K \approx 0$ of a width
$\Delta K$
yields the $h\Lambda$ residual correlation
effect at
$\langle k^*\rangle \approx p_{\rm D}m_h/(m_h+m_\Lambda)$
of a width comparable with $\Delta K$.
%
Actually, for $p_{\rm D}\ll K \ll m_i, m$,
the momentum $k^*$ is practically independent of $\zeta$:
$k^*\approx K (m/m_\Sigma)(m_h+m_\Sigma)/(m_h+m)$ and so,
for $m\approx m_h \approx m_\Sigma$ the residual
correlation function recovers the parent one for
$p_{\rm D}\ll k^* \ll m$.

As for the transfer of the $h\Sigma^0$ correlation
at small relative velocities to the $h\gamma$ correlation,
the latter is shifted to
$\langle k^*\rangle = p_{\rm D}/(1+2p_{\rm D}/m_h)^{1/2}$
with the relative width
$\Delta k^*/\langle k^* \rangle
\approx \Delta K/[\sqrt{3}\mu_{h\Sigma}(1+p_{\rm D}/m_h)] $,
where $\mu_{h\Sigma}=m_hm_\Sigma/(m_h +m_\Sigma)$ is the reduced
mass of the $h\Sigma^0$-system.
For example, if the hadron $h$ were a pion or a proton,
the residual correlation effects would be respectively situated at
$k^*\approx 34.8$ and 68 MeV/$c$ with the corresponding
relative widths $\Delta K/(332 {\rm MeV}/c)$ and
$\Delta K/(981 {\rm MeV}/c)$ much smaller than unity
provided that the parent correlation width
$\Delta K$ is less than
$\sim 100$ MeV/$c$.

\subsection{Single-channel approach}

In the following we will take the hadron $h$ to be a proton.
We thus have to calculate the FSI correlation functions for the
systems $a b = p\Lambda$ and $p\Sigma^0$
(the FSI between direct photons and protons can be neglected).
The two-particle correlation function at small $k^*$-values
is basically given by the
square of the wave function of the corresponding elastic
transition $a b \to a b$
averaged over the distance ${\bf r}^*$ of the emitters
in the two-particle c.m. system
and over the particle spin projections \cite{ll1}:
\begin{eqnarray}
\label{1}
{\cal R}({\bf p}_1,{\bf p}_2)&\doteq&
\langle |\psi_{-{\bf k}^{*}}^{S(+)}({\bf r}^{*})|^{2}
\rangle
\nonumber \\
&\doteq& 1 + \sum_S\rho_S\left[\frac12\left|\frac{f^S(k^*)}{r_0}\right|^2\right. +
\left.\frac{2\Re f^S(k^*)}{\sqrt\pi r_0}F_1(Qr_0)- \frac{\Im f^S(k^*)}{r_0}F_2(Qr_0)\right],
\end{eqnarray}
where
$F_1(z) = \int_0^z dx e^{x^2 - z^2}/z$ and $F_2(z) = (1-e^{-z^2})/z$
and $\rho_S$ is the emission probability of the two particles in a
state with the total spin $S$; we assume the emission of
unpolarized particles, i.e.
$\rho_0=1/4$ and $\rho_1=3/4$ for pairs of spin-1/2
particles.
The analytical expression in Eq. (\ref{1}) corresponds to the Gaussian
${\bf r}^*$-distribution (\ref{eq:space}).
It implies a small radius of the FSI interaction as
compared with the characteristic separation of the emitters in the
two-particle c.m. system. The non-symmetrized wave function
describing the elastic transition
can then be approximated by a superposition of the plane and
spherical waves, the latter being dominated by the s-wave,
\begin{equation}
\label{wf}
\psi_{-{\bf k}^*}^{S(+)}({\bf r}^*)\doteq
\exp(-i {\bf k}^*{\bf r}^*)+ f^S(k^*) \frac{\exp(ik^*r^*)}{r^*}.
\end{equation}
The s-wave  scattering amplitude
\begin{equation}
\label{fS}
f^S(k^*) =\frac{\eta^S\exp(2i\delta^S)-1}{2ik^*}=
(1/K^{S}-ik^*)^{-1},
\end{equation}
where $0\le \eta^S\le 1$ and $\delta^S$ are respectively
the elasticity coefficient and the phase shift,
$K^{S}$ is a function of the kinetic energy, i.e. an even function
of $k^*$. In the effective range approximation,
\begin{equation}
\label{fSa}
1/K^{S}\doteq 1/a^S +\frac12 d^S k^{*2},
\end{equation}
where $a^S$ and $d^S$ are respectively the s-wave scattering
length and effective radius at a given total spin $S$;
in difference with the traditional definition of the
two-baryon scattering length, we follow here
the same sign convention as for meson-baryon or
two-meson systems.

One can introduce the leading correction ${\cal O}(|a^S|^2d^S/r_0^3)$
to the correlation function in Eq. (\ref{1}) to account for the
deviation of the wave function (\ref{wf}) from the true solution
inside the range of the two-particle strong interaction potential
\cite{ll1}:
\begin{equation}
\label{cor1}
\Delta{\cal R}({\bf p}_1,{\bf p}_2)=
-(4\sqrt{\pi}r_0^3)^{-1}
\sum_S\rho_S |f^S(k^*)|^2 d^S(k^*) ,
\end{equation}
where the function $d^S(k^*)=2\Re d(K^S)^{-1}/d k^{*2}$;
$d^S(0)$ is the effective radius.

It should be noted that the two particles are generally
produced at non-equal times in their c.m. system and that the
wave function in Eq. (\ref{1}) should be
substituted by the Bethe-Salpeter amplitude. The latter
depends on both space (${\bf r}^*$) and time ($t^*$) separation of
the emission points in the pair rest frame and at small $|t^*|$
coincides with the wave function $\psi^S$ up to a correction
${\cal O}(|t^*/mr^{*2}|)$, where $m$ is the mass of the lighter particle.
It can be shown that the equal-time approximation in Eq. (\ref{1})
is usually valid better than
to few percent even for particles as light as pions \cite{ll1,L05}.

In this paper we use the low-energy scattering parameters for
hyperon-nucleon systems obtained within the spin-flavour SU$_6$
quark model including effective meson exchange potentials and
explicit flavour symmetry breaking of the quark Hamiltonian
to reproduce the properties of the two-nucleon system and the
low-energy hyperon-nucleon cross section data
\cite{Fujiwara}.

The $K^S$-function and the low energy scattering parameters are
real in the case of only one open channel as in the near threshold
$p\Lambda$ scattering.
For $p\Lambda$ system, we use the values from Table 6 of Ref.
\cite{Fujiwara}:
$a^0 = 2.59$ fm, $a^1 = 1.60$ fm, $d^0=2.83$ fm and $d^1=3.00$ fm.

For $p\Sigma^0$ system near threshold, there are two more open
channels, $n\Sigma^+$ and $p\Lambda$ ones, so, in principle,
one has to solve the three-channel scattering problem.
Assuming isospin conservation, this problem reduces to the
single-channel one for isospin $I=3/2$ and to the two-channel one
for isospin $I=1/2$.
The $N\Sigma(I=3/2)$ scattering parameters $a_I^S$ and $d_I^S$
are given in Table 6 of Ref.
\cite{Fujiwara}:
$a_{3/2}^0 = 2.51$ fm, $a_{3/2}^1 = -0.73$ fm, $d_{3/2}^0=4.92$ fm and $d_{3/2}^1=-1.22$ fm.
The coupling between the channel $N\Sigma(I=1/2,S=0)$ and the
$N\Lambda$ channels appears to be quite weak,
i.e. the elasticity coefficient $\eta^0(k^*)\doteq 1$,
so the low-energy scattering
parameters for the $N\Sigma(I=1/2,S=0)$ are real;
in accordance with Eq. (\ref{fS}),
the fit of the energy dependence of the $N\Sigma(I=1/2,S=0)$ phase shift
in figure 15 (fss2) of Ref. \cite{Fujiwara} yields
$a_{1/2}^0 = -1.1$ fm and $d_{1/2}^0= -1.5$ fm.
The situation is quite different for the channel $N\Sigma(I=1/2,S=1)$
which appears to be strongly coupled with the $N\Lambda$ channels
(due to the pion exchange potential)
as demonstrated by figures 15 and 31 of Ref. \cite{Fujiwara}.
As a result, the low-energy scattering parameters for the system
$N\Sigma(I=1/2,S=1)$ acquire imaginary parts:
$a_{1/2}^1 = (-1.1+i4.3)$ fm, $d_{1/2}^1= (-2.2-i2.4)$ fm.
To get these values, we have used the fact that the $K$-function
is even in $k^*$ and employed the expansions corresponding
to the effective range approximation in Eq. (\ref{fSa}):
\begin{eqnarray}
\label{even}
&&\eta(k^*)\doteq 1-2\Im a k^*+2(\Im a)^2 k^{*2}+
\left[2(\Re a)^2\Im a-2(\Im a)^3
-\Im d(\Im a)^2
\right.
\nonumber \\
&&~~~~~~~~~\left. +\Im d(\Re a)^2
+2\Re d\Re a\Im a\right]k^{*3}+
\left[-4(\Re a)(\Im a)^2+2(\Im a)^4
\right.
\nonumber \\
&&~~~~~~~~~\left.
-4\Re d\Re a(\Im a)^2+2\Im d(\Im a)^3
-2\Im d(\Re a)^2\Im a\right]k^{*4},
\nonumber \\
&&\delta(k^*)\doteq \delta(0)+\Re a k^*+
\left[-\frac43(\Re a)^3+\Re a(\Im a)^2
+\frac12\Re d(\Im a)^2
\right.
\nonumber \\
&&~~~~~~~~~\left. +\Im d\Re a\Im a
-\Re d(\Re a)^2\right]k^{*3}+
\left[-2(\Re a\Im a\left((\Re a)^2+(\Im a)^2\right)
\right.
\nonumber \\
&&~~~~~~~~~\left.
+\Re d(\Im a)^3
-\Im d(\Re a)^3
-\Re a\Im a
\left(\Im d\Im a+\Re d\Re a
\right)\right]k^{*4},
\end{eqnarray}
where $\delta(0)=0$ or $\pm \pi$.
Note however that, due to a rapid fall of the elasticity
coefficient and the phase shift near the laboratory
$\Sigma$-momentum of $\sim 100$ MeV/$c$, the use of the effective range
approximation in Eq. (\ref{fSa}) is valid up to $k^*$ of $\sim$ 50
MeV/$c$ only.

\begin{figure}[!ht]
  \begin{center}
    \includegraphics[width=16cm]{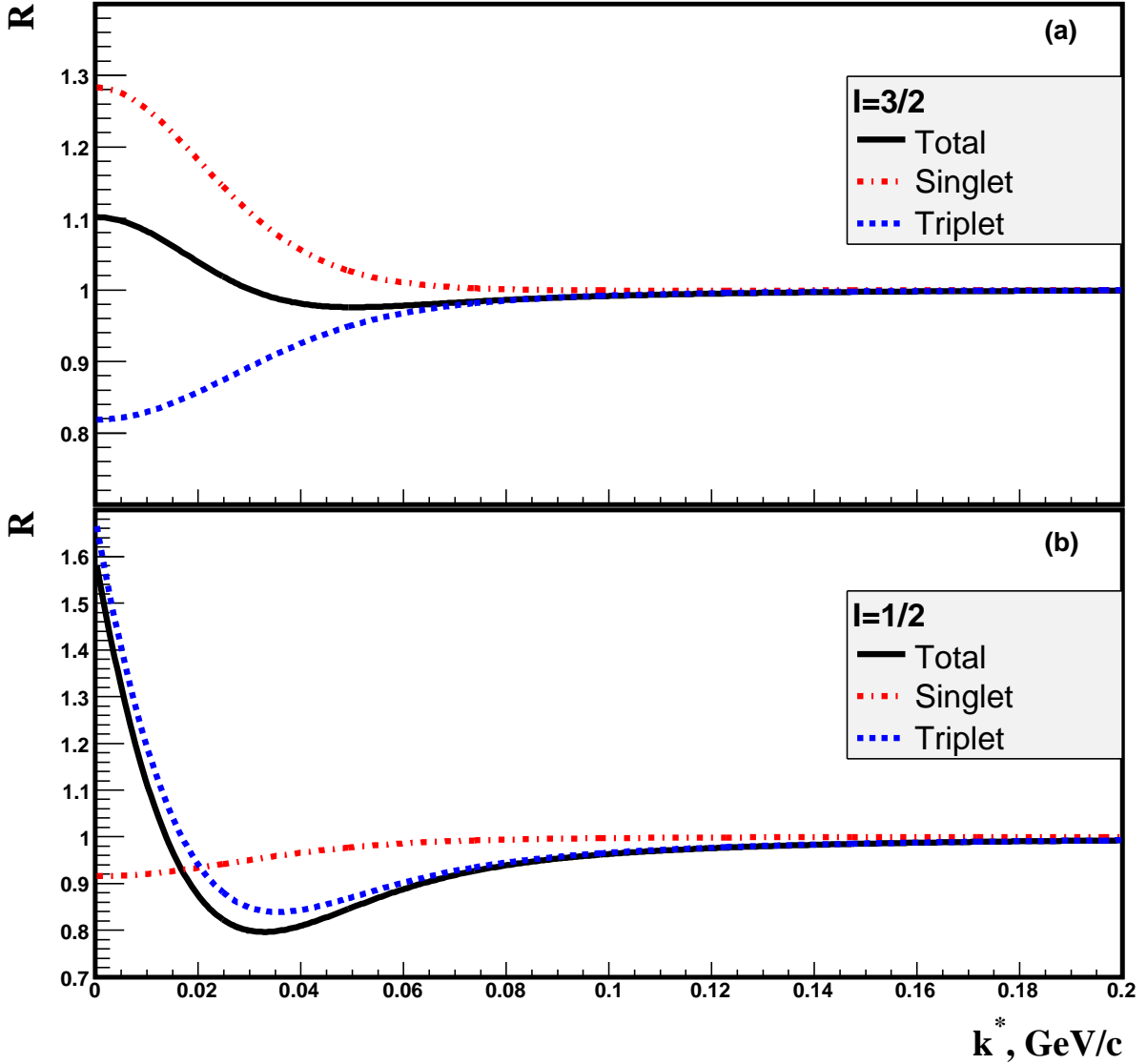}
  \end{center}
  \caption{The $N\Sigma$ correlation functions corresponding to
  isospin 3/2 (panel {\bf a}) and 1/2 (panel {\bf b}) calculated for the
  Gaussian radius $r_{0}$=3 fm assuming a uniform population of the
  spin states, i.e. $\rho_0=1/4$ and $\rho_1=3/4$. The singlet ($S=0$)
and triplet ($S=1$) contributions are shown by the
dashed-dotted and dashed curves, respectively.}
  \label{fig:NS_1o2_3j2}
\end{figure}

In Fig.~\ref{fig:NS_1o2_3j2}, we show
the $N\Sigma$ correlation functions corresponding to
isospin 3/2 (panel {\bf a}) and 1/2 (panel {\bf b}) as well as the singlet ($S=0$)
and triplet ($S=1$) contributions calculated
according to Eq. (\ref{1}) for the
Gaussian radius $r_{0}=3$ fm.
The enhancement and suppression at small  $k^*$ is related with
the positive and negative real parts of the scattering lengths,
respectively. A wide suppression of the triplet contribution to the
isospin-1/2 correlation function is due to large imaginary
parts of the corresponding scattering length and effective radius.
Though the effective range approximation is valid for this channel up
to $\sim 50$ MeV/$c$, we do not expect a substantial change
of the suppression form since at higher values of $k^*$ the
correlation function already starts to approach unity.
In any case, one may not rely on the effective range approximation
in Eq. (\ref{fSa}) and express the scattering amplitude directly
through the elasticity coefficient and the phase shift
according to Eq. (\ref{fS}).

\subsection{Two-channel approach}

The interaction of final state particles $a$ and $b$ can proceed not
only through the elastic transition $ab \to ab$ but also
through inelastic reactions of the type $cd \to ab$, where $c$
and $d$ are also final state particles of the production process. The FSI effect
on particle correlations is known to be significant
only for particles with a slow relative motion. Such particles continue to
interact with each other after leaving the domain of particle production and their
slow relative motion guarantees the possibility of the separation (factorization)
of the amplitude of a slow FSI from the amplitude of a fast production process.
For the relative motion of the particles involved in the FSI to be slow,
the sums of the particle masses in the entrance and exit channels should be close to each
other. Thus, in our case, one should account for the effect of inelastic
transition  $n\Sigma^+ \to p\Sigma^0$ in addition to the elastic
transition $p\Sigma^0  \to p\Sigma^0$.
Instead of a single channel Schr\"odinger equation one should thus solve a two-channel one
(the effect of the $p\Lambda$ channel is taken into account
in the complex effective single-channel $K_I$-functions in the isospin basis).
In solving the standard scattering problem, one should take
into account that the FSI problem corresponds to the inverse direction
of time. As a result, one has to make the substitution
${\bf k}^* (\equiv {\bf k}_a^*=-{\bf k}_b^*) \to -{\bf k}^*$ and consider
$p\Sigma^0 (\equiv 1)$ as the entrance channel and $n\Sigma^+ (\equiv 2)$
as the exit channel.
Further, in single-channel equations (\ref{wf})-(\ref{fSa}),
one has to substitute the amplitude $f^S$, the $K^S$-function, the low-energy scattering
parameters $a^S$, $d^S$ and the momentum $k^*$ by the corresponding
symmetric $2\times 2$ matrices $\hat{f}^S$, $\hat{K}^S$, $\hat{a}^S$, $\hat{d}^S$ and
$\hat{k}$:
\begin{equation}
\label{fc}
\hat{f}^S=\left[(\hat{K}^S)^{-1}-i\hat{k}\right]^{-1},~~~
(\hat{K}^S)^{-1}=(\hat{a}^S)^{-1}+\frac12\hat{d}^S k^{*2}.
\end{equation}
The momentum matrix $\hat{k}$ is diagonal
in the channel (particle) basis: $k_{ji}=k_i\delta_{ji}$;
in accordance with the energy-momentum conservation
in the transitions $1\to i$,
$k_1=k^*_a=k^*_b\equiv k^*$ and
\begin{equation}
\label{k2}
k_2=k_c^*=k_d^*=\left[2\mu_2\left(\frac{k^{*2}}{2\mu_1}+m_a+m_b-m_c-m_d\right)\right]^{1/2},
\end{equation}
where $\mu_1=m_am_b/(m_a+m_b)$ and $\mu_2=m_cm_d/(m_c+m_d)$
are the reduced masses in the channels $1=(a, b)$ and $2=(c, d)$.
Finally, the wave function $\psi^S$ in Eq. (\ref{wf}) should
be generalized to the
two-channel wave function vector $\psi^{S,i1}$ describing the
transitions $1\to i$:
\begin{equation}
\label{wfm}
\psi_{-{\bf k}^*}^{S,11}({\bf r}^*)=
\exp(-i {\bf k}^*{\bf r}^*)+ f^S_{11}(k^*) \frac{\exp(ik^*r^*)}{r^*},~~~
\psi_{-{\bf k}_i^*}^{S,21}({\bf r}^*)=
f^S_{21}(k^*)\sqrt{\frac{\mu_2}{\mu_1}} \frac{\exp(ik_2^*r^*)}{r^*},
\end{equation}
where ${\bf r}^*={\bf r}^*_a-{\bf r}^*_b$ or ${\bf r}^*_c-{\bf r}^*_d$
is the spatial separation of the particles in the exit channel.

Since the particles in both channels are members of the same isospin
multiplets, one can assume that they are produced with about the
same probability. Therefore the correlation function will be simply
a sum of the average squares of the wave functions
$\psi^{S,11}_{-{\bf k}^*}({\bf r}^*)$ and
$\psi^{S,21}_{-{\bf k}^*}({\bf r}^*)$ describing the respective elastic and
inelastic transitions \cite{LLL}. Similar to Eq. (1),
one then has:
\begin{eqnarray}
\label{1_2}
{\cal R}({\bf p}_1,{\bf p}_2)&\doteq&
\langle |\psi_{-{\bf k}^{*}}^{S,11}({\bf r}^{*})|^{2}
\rangle + \langle |\psi_{-{\bf k}^{*}}^{S,21}({\bf r}^{*})|^{2}\rangle
\nonumber \\
&\doteq& 1 + \sum_S\rho_S\left[\frac12\left|\frac{f^S_{11}(k^*)}{r_0}\right|^2\right. +
\left.\frac{2\Re f^S_{11}(k^*)}{\sqrt\pi r_0}F_1(Qr_0)- \frac{\Im f^S_{11}(k^*)}{r_0}F_2(Qr_0)\right]
\nonumber \\
&&~~+
\sum_S\rho_S\frac12\frac{\mu_2}{\mu_1}\left|\frac{f^S_{21}(k^*)}{r_0}\right|^2,
\end{eqnarray}
where the analytical expression in Eq. (\ref{1_2}) corresponds to the Gaussian
${\bf r}^*$-distribution (\ref{eq:space}); since in our case the
momentum $k_2 > k_1=k^*$ is real, the contribution of the
inelastic transition (the last term in Eq. (\ref{1_2}))
merely coincides with the quadratic term
in the contribution of the elastic transition after the substitution
$f^S_{11}\to (\mu_2/\mu_1)^{1/2}f^S_{21}$.

One should correct Eq. (\ref{1_2}) for the deviation of the spherical
waves from the true scattered waves in the inner region of the short-range
potential.
The corresponding correction $\Delta
{\cal R}$  is of comparable size to the effect of the second
channel \cite{LLL}. It is
represented in a compact form in Eq.~(125) of Ref.~\cite{L05},
similar to the single-channel correction in Eq. (\ref{cor1}). In
our case one has
\begin{eqnarray}
\label{correction}
\Delta {\cal R}({\bf p}_1,{\bf p}_2)
=
-(4\sqrt{\pi}r_0^3)^{-1}\sum_S
\rho_S\left[|f^S_{11}|^2d^S_{11}+|f^S_{21}|^2d^S_{22}\right.
+\left. 2\Re(f^S_{11}f^{S*}_{21})d^S_{21}\right],
\end{eqnarray}
where $d^S_{ij}=2\Re d(\hat{K}^S)^{-1}_{ij}/dk^{*2}$;
at $k^*=0$, $\hat{d}^S$ coincides with the real part of
the matrix of effective radii.

Assuming that the
isospin violation arises solely from the mass difference of
the particles within a given isospin multiplet,
one can express the elements of the matrices $\hat{a}^S$,
$\hat{d}^S$, $\hat{K}^S$ or $(\hat{K}^S)^{-1}$
in the channel basis through the elements
of the corresponding diagonal matrices in the
representation of total isospin $I$ (the products of the
corresponding Clebsch-Gordan coefficients being 2/3, 1/3 and
$\pm\sqrt{2}/3$). Particularly,
\begin{eqnarray}
\label{isospin}
&&({\hat K}^S)^{-1}_{11}=
\frac23 ({\hat K}^S)^{-1}_{3/2}+\frac13 ({\hat K}^S)^{-1}_{1/2}
\qquad
({\hat K}^S)^{-1}_{22}=
\frac13 ({\hat K}^S)^{-1}_{3/2}+\frac23 ({\hat K}^S)^{-1}_{1/2}
\nonumber\\
&&({\hat K}^S)^{-1}_{21}=({\hat K}^S)^{-1}_{12}=
\frac{\sqrt{2}}{3}\left[({\hat K}^S)^{-1}_{3/2}-({\hat K}^S)^{-1}_{1/2}\right].
\end{eqnarray}
Knowing the elements of the symmetric matrix $(\hat{K}^S)^{-1}$, one
can make the explicit inversion of the symmetric matrix
$(\hat{f}^S)^{-1}$ given in Eq. (\ref{fc}) and get the
required elements $f_{ij}^S$ of the
scattering amplitude matrix:
\begin{eqnarray}
\label{f-1inver}
&&D f_{11}^S=({\hat f}^S)^{-1}_{~22}=({\hat K}^S)^{-1}_{~22}-i k_2
\qquad
D f_{21}^S=-({\hat f}^S)^{-1}_{~21}=-({\hat K}^S)^{-1}_{~21}\qquad
\nonumber \\
&&D f_{22}^S=({\hat f}^S)^{-1}_{~11}=({\hat K}^S)^{-1}_{~11}-i k_1
\nonumber \\
&&D=\det(\hat{f}^S)^{-1}=({\hat f}^S)^{-1}_{~11}({\hat f}^S)^{-1}_{~22}
-[({\hat f}^S)^{-1}_{~21}]^2.
\end{eqnarray}

Note that at the momenta $k_1=k^*$ sufficiently larger than the
momentum $k_2=44.7$ MeV/$c$ of the channel $n\Sigma^+$
at the threshold of the channel $p\Sigma^0$,
one can neglect the difference between the channel momenta
and apply the relations (\ref{isospin}) directly to
the elements of the amplitude matrix $\hat{f}^S$.

\subsection{Results}

\begin{figure}[!h]
  \begin{center}
    \includegraphics[width=18cm]{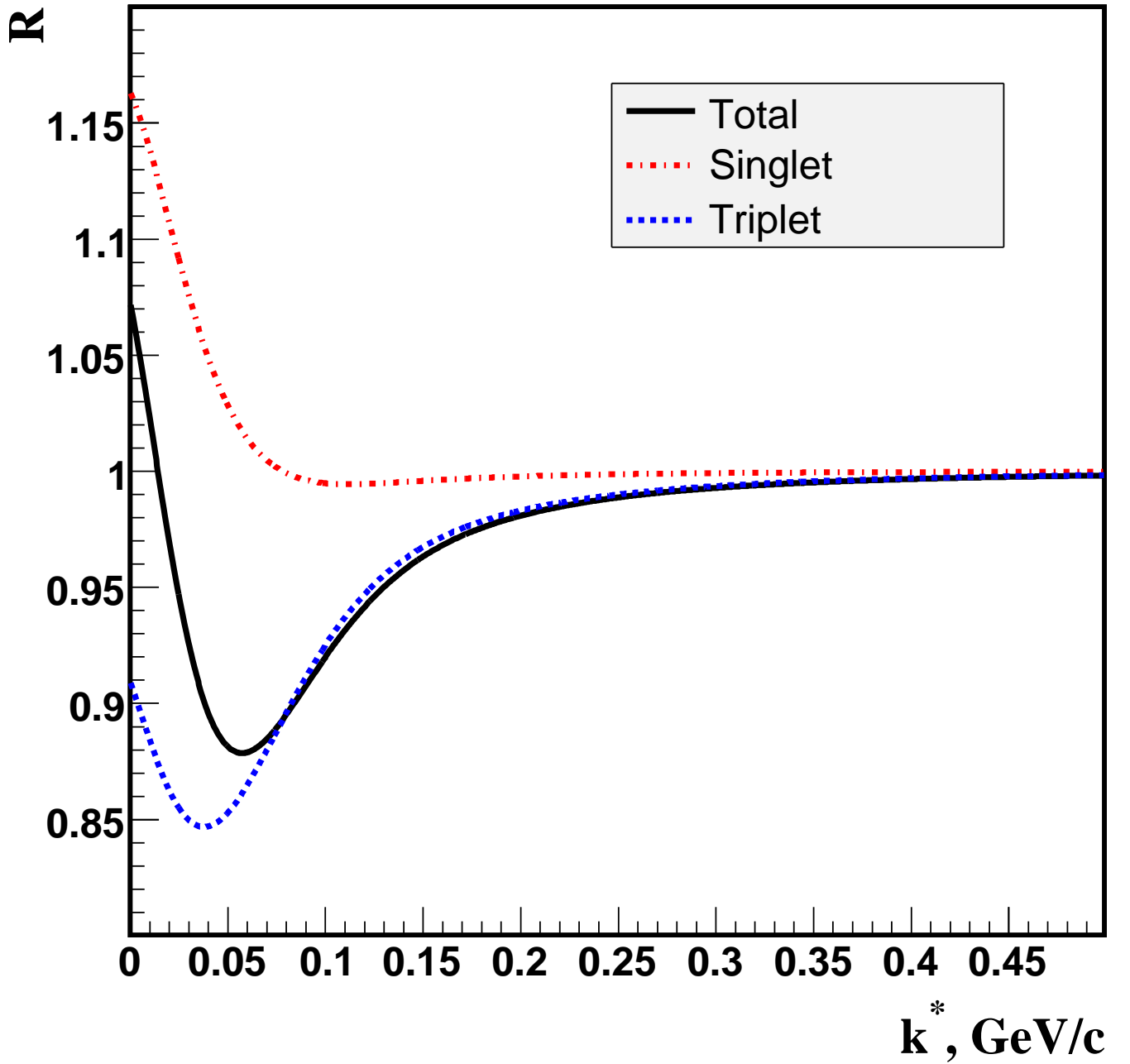}
  \end{center}
  \caption{The $p\Sigma^0$ correlation function calculated for the
  Gaussian radius $r_{0}$=3 fm assuming a uniform population of the
  spin states.
  The singlet ($S=0$)
  and triplet ($S=1$) contributions are shown by the
  dashed-dotted and dashed curves, respectively.}
  \label{fig:pSigma0}
\end{figure}

The $p\Sigma^0$ correlation function as well as the singlet ($S=0$)
and triplet ($S=1$) contributions calculated for the
Gaussian radius $r_{0}=3$ fm are shown in Fig.~\ref{fig:pSigma0}.
As already mentioned in the discussion of
Fig.~\ref{fig:NS_1o2_3j2},
the enhancements at small  $k^*$ are related with
the positive real parts of the scattering lengths and
a wide suppression of the triplet contribution is due to
large imaginary
parts of the isospin-1/2 scattering length and effective radius.

\begin{figure}[!h]
  \begin{center}
    \includegraphics[width=16cm]{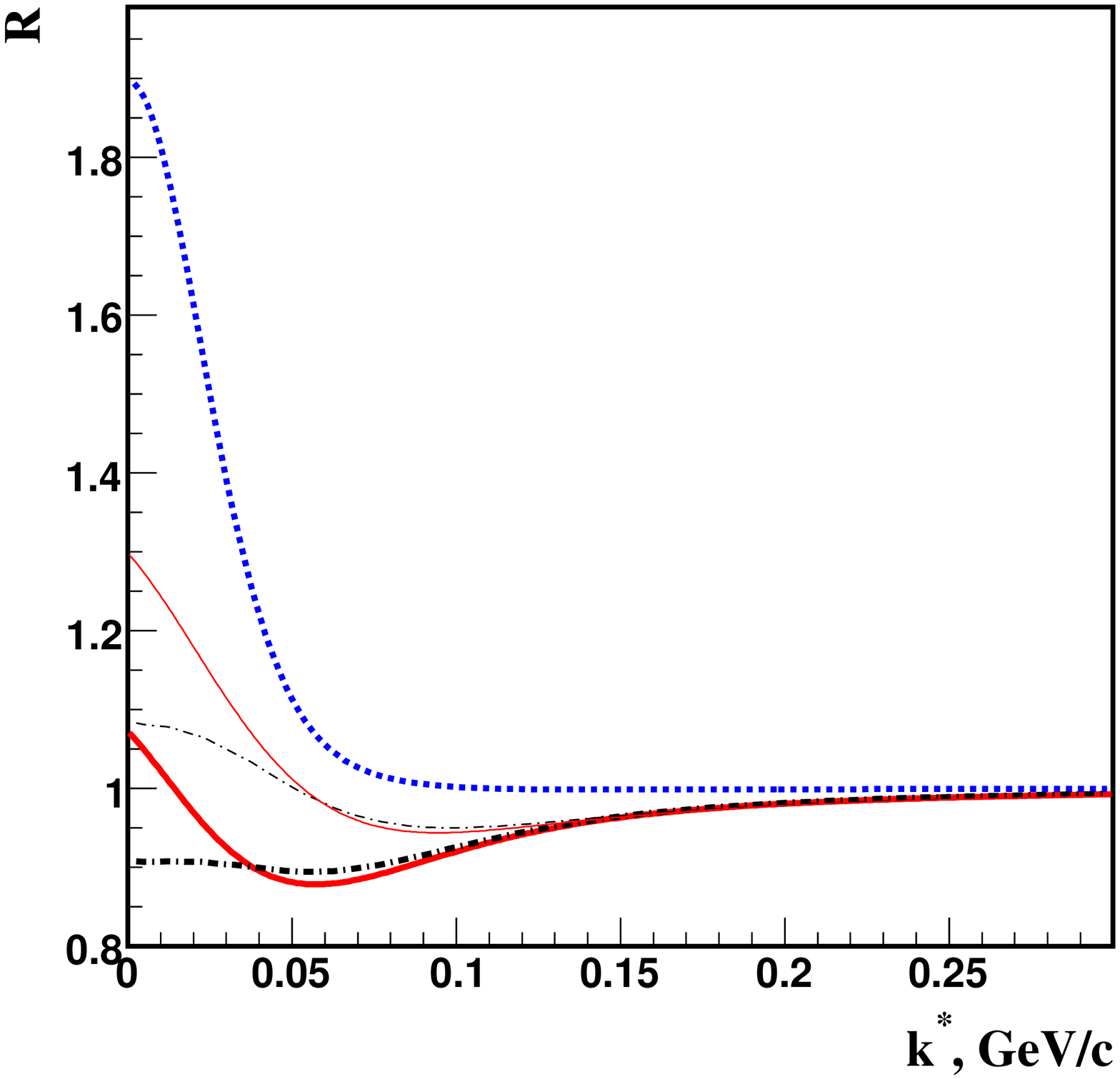}
  \end{center}
  \caption{The $p\Sigma^0$ (solid curve), the $p\Lambda$ (dashed curve)
  and the residual $p \Lambda_{\Sigma^{0}}$ (dashed-dotted curve)
  correlation functions calculated for the
  Gaussian radius $r_{0}$=3 fm assuming a uniform population of the
  spin states and the thermal-like momentum distribution (\ref{eq:momentum})
  with $T_0=168$ MeV/$c$.
  The thin solid and thin dashed-dotted curves correspond to
  the $p\Sigma^0$ and the residual $p \Lambda_{\Sigma^{0}}$
  correlation functions calculated with
  the scattering parameters
  $a_{1/2}^1 = (2.54 + i 0.26)$ fm, $d_{1/2}^1 = 0$
  obtained from the pole position in the NSC89
  model \cite{miy99} on the assumption of vanishing effective radius.
}
  \label{fig:pSigma0pLambda}
\end{figure}

In Fig. \ref{fig:pSigma0pLambda}, we compare
the $p\Lambda$ correlation function
with the $p\Sigma^0$ and the residual $p\Lambda_{\Sigma^0}$
ones calculated at the same conditions.
Note that in our model the parent correlation functions are
independent of the single particle spectra contrary to the residual
correlations. To calculate the latter, we have used the
thermal-like distribution (\ref{eq:momentum}) with $T_0=168$ MeV/$c$.
One may see that the $p\Lambda_{\Sigma^0}$ residual correlation
function is quite different from the $p\Lambda$ one.
In accordance with the discussion after Eq. (\ref{k_k}), the
former is close to the parent $p\Sigma^0$ correlation function for
$k^* > 70$ Mev/$c$.
In high energy heavy ion collisions
the fraction of $\Lambda$'s from  $\Sigma^0$ decay
is $\sim 40\%$. If the corresponding residual correlation
were neglected, the parent $p\Lambda$ correlation effect would be
underestimated thus leading to an overestimation of the source size.

\begin{figure}[!t]
  \begin{center}
    \includegraphics[width=16cm]{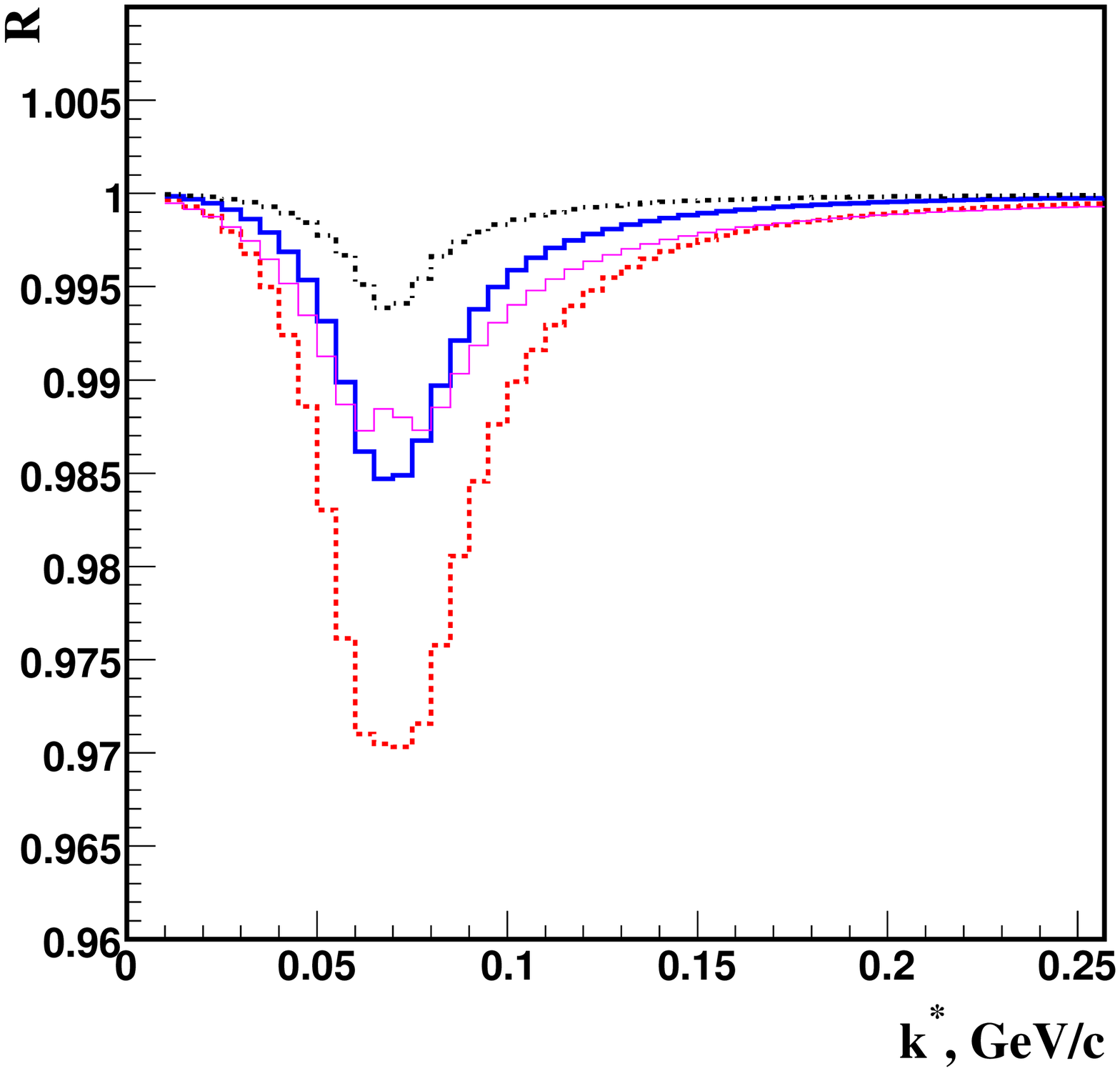}
  \end{center}
  \caption{The residual $p \gamma_{\Sigma^{0}}$ correlation function
  resulting from the parent $p \Sigma^{0}$ correlation due to the
  $\Sigma^{0}\to \Lambda\gamma$ decay calculated for the Gaussian
  source radius $r_0 = 2$ fm (dashed curve), $r_0 = 3$ fm (solid curve)
  and $r_0 = 5$ fm (dashed-dotted curve). The thermal-like momentum distribution
  (\ref{eq:momentum}) with $T_0=168$ MeV/$c$ is assumed for parent particles.
  The thin solid curve corresponds to $r_0 = 3$ fm and
  the scattering parameters
  $a_{1/2}^1 = (2.54 + i 0.26)$ fm, $d_{1/2}^1 = 0$
  obtained from the pole position in the NSC89
  model \cite{miy99} on the assumption of vanishing effective radius.
}
  \label{fig:pgamma}
\end{figure}

The residual $p\gamma$ correlation functions resulting
from the parent $p\Sigma^0$ correlation due to the
$\Sigma^0\to\Lambda\gamma$ decay
calculated for different Gaussian radii of the source
are shown in Fig.~\ref{fig:pgamma}.
In accordance with the discussion after Eq. (\ref{k_k}),
the parent correlation at small relative velocities is
shifted to rather narrow $k^*$-region centered at $\sim 70$
MeV/$c$. Fig.~\ref{fig:pgamma} also demonstrates the sensitivity
of the residual correlation effect to the source size.

It should be noted that there exists substantial uncertainty
in the theoretical predictions for the low-energy scattering
parameters in the isospin-1/2 $N\Sigma $-channel.
Thus the predictions of various Nijmegen potential models
for the near-threshold pole position $\alpha_{1/2}^1$
in this channel \cite{miy99} yield
in the limit of zero effective radius,
when $a_{1/2}^1= \alpha_{1/2}^{1*}/|\alpha_{1/2}^1|^2$,
similar triplet scattering length $a_{1/2}^1$
as that deduced from Ref. \cite{Fujiwara} in the case
of NSC97f and NF potentials while, they yield even opposite
sign of the real part of this scattering length
in the case of earlier NSC89 and ND potentials
\cite{kerbikov}.
To demonstrate the effect of possible uncertainty, we present
in figures \ref{fig:pSigma0pLambda} and \ref{fig:pgamma},
besides the correlation functions corresponding to the
potential model of Ref. \cite{Fujiwara}, also those obtained
from the pole position in the NSC89 model assuming
$d_{1/2}^1 = 0$ \cite{kerbikov}:
$a_{1/2}^1 = (2.54 + i 0.26)$ fm.
One may conclude from these figures that
the  shape and the intensity of the $p \Lambda_{\Sigma^{0}}$
and $p \gamma_{\Sigma^{0}}$ residual correlations
are sensitive to the $p\Sigma^0$  FSI and
source  size  parameters thus providing a new
possibility to learn about these parameters.

The fraction of residual $p\gamma$ correlations arising from
parent $p\Sigma^0$ correlations
with respect to all other contributions into $p\gamma$
system is not so large as for the $p\Lambda$ system.
The background arises mainly from photons from $\pi^0$
decay. Such photons
can reduce the effect of our interest in $p\gamma$ system and
make it invisible. The
methods of the background suppression depend
on experimental details and should be discussed
separately.

\section{Conclusion}

The two-photon and proton-photon residual correlations
can serve as a new important source
of information on the FSI and/or source size parameters
as well as on the direct particle fractions.
Particularly, a nontrivial femtoscopic irregularity
in the proton-photon correlation function centered
at $k^*\approx 70$ MeV/$c$
is expected due to the $p\gamma_{\Sigma^0}$
residual correlation.
It is shown that the $p\Lambda_{\Sigma^0}$ residual correlation
recovers the negative parent $p\Sigma^0$ correlation function for
$k^* > 70$ Mev/$c$. The neglect of this negative
residual correlation would lead to the underestimation of
the parent $p\Lambda$ correlation effect and
to an overestimation of the source size.

\bigskip
\noindent\textbf{Acknowledgements}
This work was supported by the
RosAtom,
the Grant of the Russian Foundation for Basic Research
under Contract No. 04-02-17468a and 06-08-01555a, the
Grant Agency of the Czech Republic under contract 202/07/0079
and partly carried out within the scope of the GDRE:
Heavy ions at ultrarelativistic energies -–
a European Research Group comprising
IN2P3/CNRS, EMN, University of Nantes,
Warsaw University of Technology, JINR Dubna, ITEP Moscow and
BITP Kiev.



\pagestyle{empty}

\end{document}